\documentclass[english,numbers,authoryear]{elsarticle}
\usepackage[T1]{fontenc}
\usepackage[latin9]{inputenc}
\usepackage{array}
\usepackage{booktabs}
\usepackage{graphicx}
\usepackage{esint}

\makeatletter

\providecommand{\tabularnewline}{\\}

\newcommand{\apj}{ApJ}
\newcommand{\apjs}{ApJS}
\newcommand{\apjl}{ApJL}

\newcommand{\aap}{A~\&~A}

\newcommand{\mnras}{MNRAS}
\newcommand{\solphys}{Sol.Phys.}

\newcommand{\araa}{An. Rev. Astron. Astroph.}
\newcommand{\ssr}{Space Sci. Rev.}
\newcommand{\pasp}{Publications of the Astron. Soc. of Pacific}
\newcommand{\nat}{Nature}

\usepackage{babel}

\usepackage{babel}

\usepackage{babel}

\usepackage{babel}

\makeatother

\usepackage{babel}
\begin{document}
\begin{frontmatter}

\title{Large-scale magnetic fields and anomalies of chemical composition
of stellar coronae }

\author{V.V. Pipin and V.M. Tomozov }

\address{Institute Solar-Terrestrial Physics, Irkutsk}
\begin{abstract}
We present evidence that anomalies in abundance of the chemical admixtures
with the low first ionization potential (FIP) in the low corona of
the late-type stars can be related to a topology of the large-scale
magnetic field. Observational data from Ulysses and the Stanford Solar
Observatory reveal the high correlations between the FIP effect of
the solar wind and amount of the open magnetic flux. To analyze the
stellar activity data we relate the amount of the open magnetic flux
with the ratio between poloidal and toroidal magnetic field of a star.
The solar-type stars show the increase abundance of the low FIP elements
relative to elements with the high FIP with the decrease of the large-scale
poloidal magnetic field (and increase the toroidal component of the
magnetic field). The branch of the fully convective stars demonstrates
inversion of the FIP-effect. This inversion can result from strong
coronal activity, which is followed by the strong poloidal magnetic
on these stars. 
\end{abstract}
\end{frontmatter}

\pagebreak{}

\section{Introduction }

Observations show that the stellar magnetic activity depends largely
on the stellar mass and the rotational period (see, \citealp{2009ARAA_donat,vid14MNR,2016MNRAS1129S}).
These parameters determine a level of the magnetic activity (\citealp{1984ApJ...287..769N})
and the typical topology of the large-scale magnetic field (\citealt{2008ASPC..384..156D,2016MNRAS1129S}).
Therefore, we can, generally, expect that the chromospheric and coronal
activity in late-type stars varies with the spectral class and rotational
period, correspondingly (\citealp{2012LRSP9}).

It is widely accepted that chemical composition in the solar photosphere
is spatially uniform. On another hand, \citet{1963MNRAS125.543P}
showed that the coronal chemical composition can depend on the first
ionization potential (hereafter FIP). Moreover, elements with low
FIP ($\le10$ eV), for example, Fe, Mg, Si, K etc, show an increase
of abundances in the coronal loops above the active regions. At the
same time, elements with high FIP (FIP > 10 eV) show no variation
or, sometimes, even a little-decreased composition there. This is
so-called 'FIP-effect'. General information about manifestations of
FIP effect in solar atmosphere, solar wind and in composition of solar
energetic particles are presented in Meyer's review \citeyearpar{1985ApJS57.151M}.

Observations showed that elements with low FIP can be accumulated
at top of the closed magnetic loops in course of the active regions
evolution. For example, \citet{2001ApJ555.426W} found that chemical
composition of the coronal plasma above recently emerging active region
is close to the composition at the photosphere, and after that, it
shows increased concentrations of the elements with low FIP at the
top of a coronal loop. The ratio between the coronal and photospheric
abundances of low FIP elements can reach factor 4 after few days of
the active region evolution. Also, they found that in the long-lived
active regions the increasing concentration of the elements with low
FIP can reach factor 8 or even higher than that value. Observations
show complicated distributions of the low FIP admixtures in the coronal
plasma above the active regions. Recently, new interesting results
have been obtained on FIP effect evolution and distribution in solar
active structures. Using results of observations of EIS spectrometer
onboard Hinode, \citet{2015ApJ802.104B} found that the low FIP element
enrichment occurs in upper parts of coronal loops located in central
zones of the active region. At the active region periphery, a process
of magnetic reconnection with emerging small-scale magnetic fields
with photospheric plasma composition neutralizes accumulation of the
low FIP elements. As a result, the mixed plasma composition with reduced
FIP bias is transferred by interchange reconnection beyond active
region boundaries and can manifest itself in slow wind composition. 

The detailed observations have shown that element composition in a
hot plasma, which evaporated during large X-ray flares turned out
to be close to the composition of the photosphere \citep{2014ApJ786L.2W}.
At the same time, plasma composition in CME events connected with
of such flares was heavily enriched of low FIP elements with a high
ionization degree \citep{Zurbuchen2016}. Additionally, with EIS spectrometer
on Hinode measurements was made of the plasma composition during impulsive
heating events in the solar transition region. During these events,
spectral lines of high FIP elements were amplified with respect to
low FIP lines and showed a composition close to photospheric one.
On the contrary, overlying long-lived coronal fan loops showed low
FIP element enrichment \citep{2016ApJ824.56W}. These authors concluded
that plasma composition is an important feature of the coronal heating
process, with impulsive heating (such as a flare or coronal ejection)
leading to evaporation of photospheric plasma; and in higher frequency
heating (for example, wave heating) resulting in enrichment low FIP
in the long-lived coronal structures.

The above consideration shows that the FIP fractionation of the solar
coronal plasma is a rather complicated process. The key questions
for us are as follows. Firstly, is the FIP fractionation controlled
by the global parameters of the solar activity? Secondly, what are
those global parameters of the magnetic activity that determine the
magnitude of the FIP-effect for the Sun as a star? These questions
are motivated by the solar and stellar observations. Using observational
data from SDO Brooks et al. (2017) showed that variations of the integral
FIP-effect for the Sun as a star is correlated with a proxy for solar
activity, the F10.7 radio flux, and therefore, with the solar cycle
phase. Exploring the late-type stars, \citet{2012ApJ753.76W} found
a systematic decrease of the FIP effect with the stellar mass. Moreover,
their results show an inversion of the FIP-effect for the low mass
M-dwarfs (inversed FIP effect). Results of \citet{2012ApJ753.76W}
were confirmed by \citet{Brooketal17fip}. The global parameters of
the stellar magnetic activity changes with the stellar mass (Donati
\& Landstreet 2009). From earlier studies (see, e.g., \citealt{skum72,1984ApJ...287..769N})
we know that another critical parameter for stellar magnetic activity
is the stellar rotation rate.

Interesting that the results of \citet{2012ApJ753.76W} do not show
a relation between the coronal FIP effect and coronal activity parameters
such as the X-ray luminosity ($\log L_{X}$). The same was found for
the surface X-ray flux ($\log F{}_{X}$). On another hand, these parameters
show clear correlations with rotation rate and the Rossby number of
a star (see, \citealt{2003ApJ598.1387P,2016Natur535.526W}). This
motivate us to explore the connection of the coronal chemical composition
with other parameters of the stellar magnetic activity. One of the
most important global parameter of the stellar coronal field is an
amount of the open and closed magnetic flux. This parameter is tightly
related with the strength of the stellar wind and the magnetic braking
of the angular momentum of a star \citep{1968MNRAS138.359M}. 

\citet{2015SSRv188.211F} pointed out that chemical composition of
plasma of the slow component of the solar wind shows an increase concentration
of elements with low FIP by factor 3 compares to their values for
the fast component of the solar wind. \citet{2004ApJ612.1171W} showed
that the so-called ``slow'' and ``fast'' components of the solar
wind can be related to the closed and open magnetic flux of the Sun.
This is employed in the solar wind models \citep{2012SSRv172.123W}.
Note, that in the global models of the solar coronal field the definition
of the open and closed magnetic structure is phenomenological and
it is related to the so-called ``source surface'' (\citealt{1969SoPh6.442S}).
\citet{2004ApJ614.1063L} employ the magnetic topology to model the
solar coronal FIP effect. In his model, it is assumed that FIP fractionation
is induced by the so-called ponderomotive force which results from
propagation of the resonant alfvenic waves along the closed magnetic
field lines. Direction and amplitude of the ponderomotive force depend
on the gradient of wave energy in the low corona. For the resonant
waves the energy grows up and this results in the upward ponderomotive
force acting on ions of the low FIP. In the open magnetic field structures
and for the off-resonant waves the wave energy decreases in the major
part of the low corona. In following to ideas of \citet{2004ApJ612.1171W}
and \citet{2004ApJ614.1063L} let us assume that the average FIP effect
of the Sun as a star approximately reflects the ratio between amounts
of the open and closed fluxes of the solar coronal magnetic field.
{In our analysis we have to keep in mind that the open field
configurations can have a slow wind composition and closed field can
have a photospheric composition. This may be a source of systematic
error in our study.}

The purpose of the paper to study the statistical relationships between
the global parameters of the stellar magnetic activity such as of
the open and closed magnetic flux and the integral parameter of the
coronal FIP effect. Both the solar and stellar activity observations
data will be discussed. In our study, we use a combination of data
from results of Wood et al. (2012) and other data sets including the
solar wind composition from the Ulysses space mission, the magnetic
field observations from the Wilcox Solar Observatory, the stellar
magnetic field spectropolarimetric measurements (see, Petit et al.,
2014 and Marsden et al., 2014). In next section, we will describe
the datasets, which will be used in this paper.

\section{Data}

\subsection{Solar observations}

To trace solar cycle variations of the FIP effect of the Sun as a
star we use observations of Ulysses spacescraft. The data from instrument
Ulysses/SWICS ( Solar Wind Ion Composition Spectrometer) are used.
It is assumed that in-situ measurements of the wind's composition
of heavy ions, e.g., such as O, C, and Fe, can be used to infer its
composition in the low corona \citep{1986SoPh103.347B}. In this paper
we will use the SWICS's measurements for ions of Fe and O. The measurements
procedure and their implications were described in details by \citet{1995SSRv72.49G}
and \citet{1995SSRv72.71V}. The observed ratio (Fe/O) gives a proxy
for the magnitude of the FIP fractionation in the solar corona. {The
magnitude of the FIP effect is defined as follows:} 
\begin{equation}
\mathrm{FIP}=\frac{\left[Fe/O\right]_{c}}{\left[Fe/O\right]_{ph}}.\label{eq:1}
\end{equation}
Following to procedure given by \citet{2012ApJ753.76W} and \citet{2015LRSP12.2L}
we introduce the so-called ``FIP bias''. It is determined by ratio
of the photospheric value of Fe/O and its coronal counterpart, i.e,
\begin{equation}
\mathrm{FIP_{bias}}=\log\left[Fe/O\right]_{ph}-\log\left[Fe/O\right]_{c}.\label{eq:fip}
\end{equation}
Therefore, we have $\mathrm{FIP_{bias}}=-\log\mathrm{FIP}$. Hereafter,
we assume that $\log\left[Fe/O\right]_{ph}=0.06$ \citep{2015LRSP12.2L}.
The relative error of the SWICS's measurements of the elemental composition
is estimated to be 10- 25\% (see, e.g., \citealp{1995SSRv72.49G}).

\begin{figure}
\includegraphics[width=0.9\columnwidth]{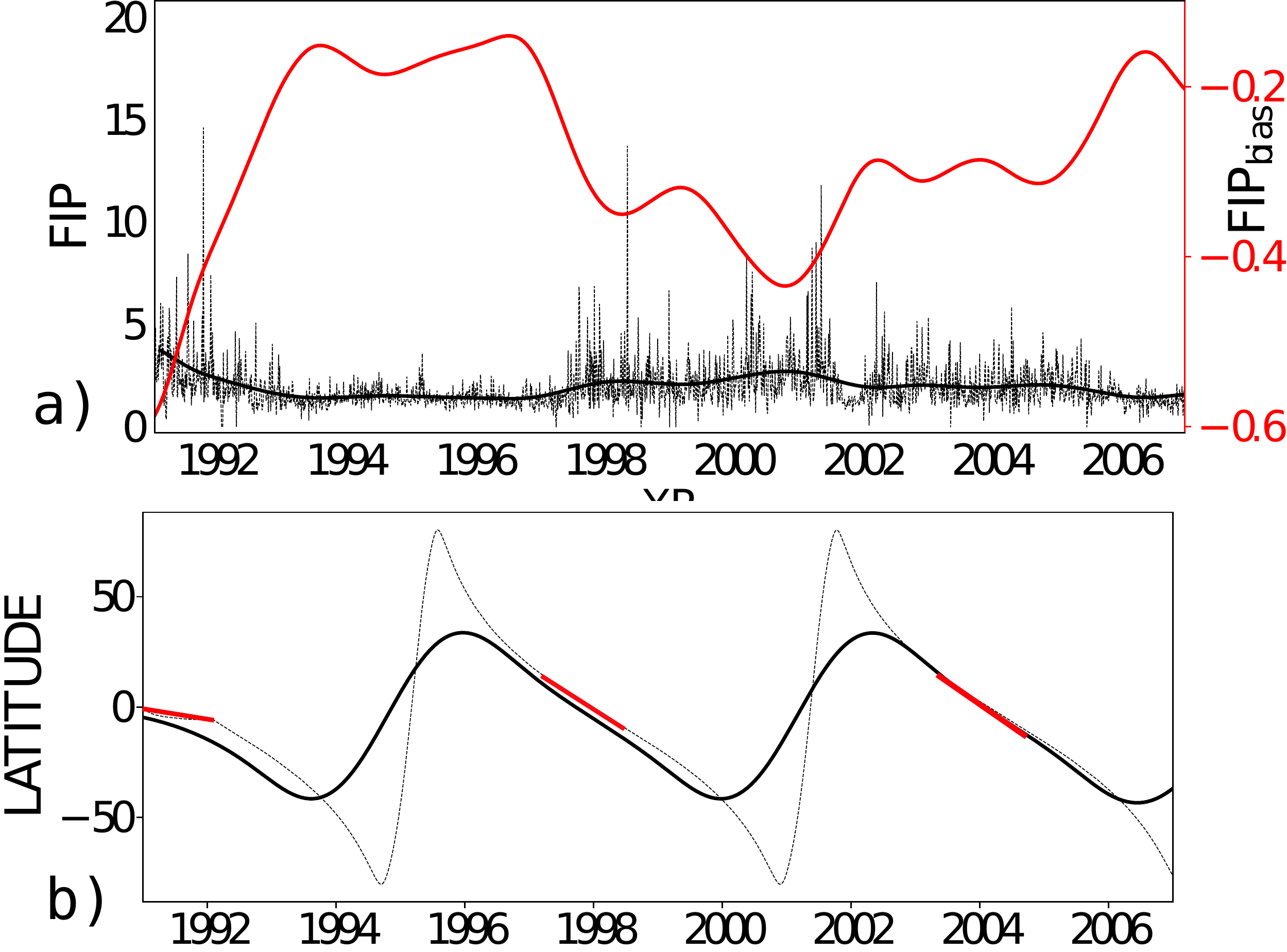}

\caption{\label{fig:sun}a) The FIP fractionation (see, Eq.1) from the Ulysses/SWICS,
the original data (dashed curve, 3-days cadence) and after the Gaussian
filtering (the solid curve). The red line curves show the $\mathrm{FIP_{bias}}$
(see, Eq.2) with magnitude shown by the right y-axis; b) Variations
of the helio-latitude of the ULYSSES spacecraft (dotted curve). The
solid curve shows the ``effective\textquotedblright{} latitude of
the observations after smoothing. The red patches show the separated
intervals for analysis (see, the main text).}
\end{figure}

Superposed epoch analysis of \citealp{1995SSRv72.49G} showed the
rather tight correlation between the magnitude of FIP fractionation
and coronal structures related to regions of the fast and slow solar
wind. Their results clearly showed that FIP fractionation is small
in coronal holes. These solar phenomena are usually related to the
open magnetic field structures. On another hand, the FIP fractionation
grows by factor 3-4 above the active regions, where the closed magnetic
field structures dominate. This conclusion was further stressed by
\citet{2015SSRv188.211F}.

The trajectory of the Ulysses spacecraft passed over very different
parts of the solar disk. To deduce the FIP effect for the Sun as a
star we smoothed the data by the Gaussian filter with the FWHM equal
one year. The result is shown in Figure \ref{fig:sun}. The graph
shows that in the smoothed data the FIP fractionation had a maximum
magnitude of factor 4. It occurs near the maximum of the 22 solar
cycle. The corresponded value of FIP bias was about -0.6. \citet{2015LRSP12.2L}
and \citet{Brooketal17fip} argued that the maximum FIP bias for the
Sun as a star during epochs of solar maxims should be around $-0.6$.
However, their results are related to cycles 20 and 24. Those cycles
have the considerable smaller magnitudes than cycles 22 and 23. For
the minimum of cycle 23 \citet{Brooketal17fip} found $\mathrm{FIP_{bias}}\approx-0.16$.
This roughly agrees with our estimation from the smoothed data.

The aim of the smoothing procedure is to get an estimation of the
FIP effect of the Sun as a star. Another purpose is to eliminate the
possible correlation between the measured Fe/O ratio and the spacecraft
latitude (see Figures \ref{fig:sun}a and b). However, we must take
into account the fact that the latitude of the spacecraft's flight
around the Sun varied non-monotonically, and, in some periods, the
trajectory passed above equatorial latitudes during a long time (more
than one year). As a result, the applied smoothing procedure may yield
a distorted estimate of the Sun as a star's FIP effect, in spite of
averaging over a large number of solar rotation. To analyze this in
detail we look at those particular periods of times. The data sets
include the following time intervals: 1991-1992, 1997-1999 and 2003-2005.
The corresponded patches are marked by red color, see Figure\ref{fig:sun}b.
For these periods of time the original Ulysses data were averaged
on the interval of the one Carrington rotation. In this case, the
averaged FIP effect represent contributions from the equatorial part
of the Sun. 

Parameters of the solar magnetic field were determined using observations
of the Wilcox Solar Observatory (WSO), see \citet{WSO1} and \citet{WSO2}.
These data span about four solar cycles starting from 1976. The large-scale
magnetic field was restored using the spherical harmonic coefficients
obtained from the potential field extrapolation employing the so-called
``source surface'' at 2.5R. More details about the procedure to
determine the spherical harmonic coefficients decomposition can be
found at the WSO website. Using the obtained distributions of the
large-scale magnetic field we determine the unsigned open, $\Phi_{o}\left(t\right)$,
the closed , $\Phi_{c}\left(t\right)$, and the total, $\Phi_{s}\left(t\right)$,
magnetic fluxes as follows 
\begin{eqnarray}
\Phi_{o}\left(t\right) & = & R_{s}^{2}\int\left|B_{r}\left(\theta,\phi,t\right)\right|_{r=R_{s}}\sin\theta d\phi d\theta,\\
\Phi_{s}\left(t\right) & = & R^{2}\int\left|B_{r}\left(\theta,\phi,t\right)\right|_{r=R}\sin\theta d\phi d\theta,\\
\Phi_{c}\left(t\right) & = & \Phi_{s}\left(t\right)-\Phi_{o}\left(t\right)
\end{eqnarray}
where $\phi$ is longitude, $\theta$ is the polar angle and the source
surface radius is $R_{s}=2.5R$. Also we need a proxy for the ratio
$\Phi_{o}\left(t\right)/\left(\Phi_{s}\left(t\right)\right)$. For
this we suggest to employ the mean magnitudes of the \emph{axisymmetric
} poloidal and total (axisymmetric and nonaxisymmetric) toroidal magnetic
field, which are as follows, $\overline{B^{(PA)}}=\left\langle \sqrt{B_{r}^{2}\left(\theta,t\right)+B_{\theta}^{2}\left(\theta,t\right)}\right\rangle _{r=R}$,
and $\overline{B^{(T)}}=\overline{\left|B_{\phi}\left(\theta,\phi,t\right)\right|}_{r=R}$,
where the magnetic field components are inferred from the same spherical
harmonic coefficients. We introduce another parameter as a proxy for
the ratio $\Phi_{o}\left(t\right)/\left(\Phi_{s}\left(t\right)\right)$:
\begin{equation}
P_{AX}=\frac{\overline{B^{(PA)}}}{\overline{B^{(PA)}}+\overline{B^{(T)}}}.\label{eq:pax}
\end{equation}

\begin{figure}
\includegraphics[width=0.9\columnwidth]{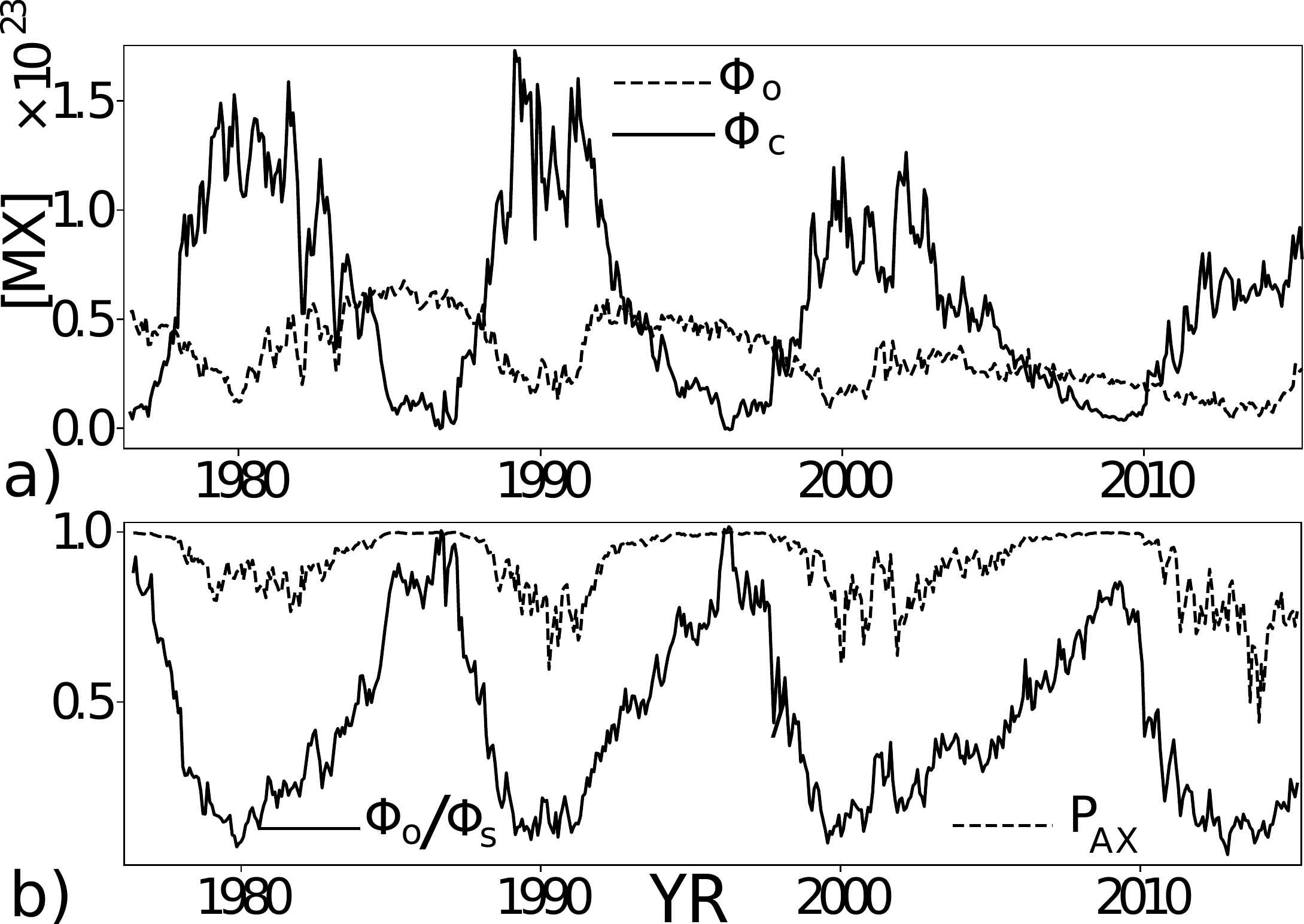}

\caption{\label{fig:WSO}a) The unsigned open and closed magnetic fluxes deduced
from the WSO's harmonic coefficients; b) ratio $\Phi_{o}\left(t\right)/\left(\Phi_{s}\left(t\right)\right)$
(solid black line); the dashed line shows the parameter $\mathrm{P_{AX}}$
(the relative contribution of the axisymmetric poloidal field, see
Eq\ref{eq:pax}). {The points on the graphs represent the individual
Carrington rotations.}}
\end{figure}

\begin{figure}
\includegraphics[width=0.98\columnwidth]{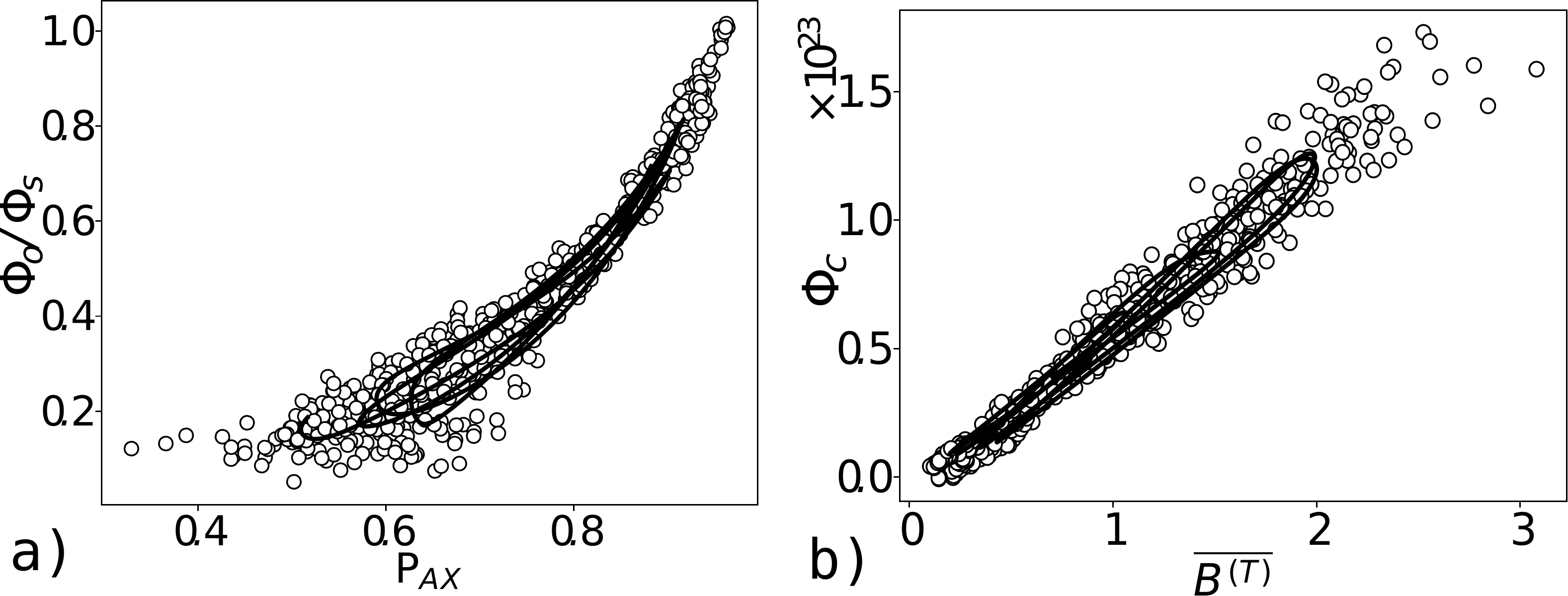}

\caption{\label{figmag}a) Relation between parameters $\mathrm{P_{AX}}$ and
$\Phi_{o}/\Phi_{s}$ for the individual rotations (circles) and the
smoothed time series (solid line); b) the same as (a) for $\Phi_{c}$
vs $\overline{B^{(T)}}$.}
\end{figure}

The reason to express the relative contribution of the axisymmetric
poloidal magnetic field in form of Eq(\ref{eq:pax}) is that it correlates
well with $\Phi_{o}/\Phi_{s}$. Figure \ref{fig:WSO} shows results
for the defined parameters of the solar magnetic field. {The points
on the graph represent the individual Carrington rotations. The given
time profiles of the parameters $\Phi_{o}\left(t\right)/\left(\Phi_{s}\left(t\right)\right)$,
$\Phi_{c}\left(t\right)$, $\mathrm{P_{AX}}$ and $\overline{B^{(T)}}$
were further smoothed} with the same Gaussian filter as the data
about chemical composition. {Note, that the $\overline{B^{(T)}}$
represent the non-axisymmetric toroidal magnetic field on the solar
surface. }

{We have no estimations for the total magnetic flux parameters for
our sample of stars. For this purpose, we will employ $\mathrm{P_{AX}}$
( the relative contribution of the axisymmetric poloidal field) as
a proxy of $\Phi_{o}/\Phi_{s}$, and the parameter $\overline{B^{(T)}}$
(the mean strength of the nonaxisymmetric toroidal magnetic field)
is a proxy for the unsigned total closed flux $\Phi_{c}$. Figures
\ref{figmag}a and b show that our choice is reasonable. In both cases,
we found the tight relations in point to point correlations and in
the smoothed time-series, as well. The relation of $\Phi_{o}/\Phi_{s}$
and $\mathrm{P_{AX}}$ seems to be a power law. The best approximation
is as follows: $\Phi_{o}/\Phi_{s}\sim\mathrm{P_{AX}}^{4}$. On another
hand, $\Phi_{c}$ and $\overline{B^{(T)}}$ are linearly connected.}

{Parameters of the linear regressions for $\Phi_{o}/\Phi_{s}\sim\mathrm{P_{AX}}^{4}$
and$\Phi_{c}\sim\overline{B^{(T)}}$ are given in Table\ref{sper-1}.
In both cases the correlation coefficients are about $0.98$. Estimations
of the Spearman's rank correlations give the similar correlations
coefficients, i.e., r=0.975 for $\Phi_{o}/\Phi_{s}\sim\mathrm{P_{AX}}^{4}$
and r=0.967 for $\Phi_{c}\sim\overline{B^{(T)}}$. Using these results
we will try to connect the stellar FIP$_{bias}$ observation with
magnetic parameters $\mathrm{P_{AX}}$ and $\overline{B^{(T)}}$.
}

\begin{table}
\begin{centering}
\begin{tabular}{ccccc}
\toprule 
 & s  & p  & r  & $\sigma$\tabularnewline
\midrule 
{$\Phi_{o}/\Phi_{s}\sim\mathrm{P_{AX}}^{4}$}  & 1.02  & 0.0  & 0.98  & 0.01\tabularnewline
\midrule 
{$\Phi_{c}\sim\overline{B^{(T)}}$}  & 6.44$\cdot10^{22}$$\mathrm{\frac{Mx}{G}}$  & 0.0  & 0.97  & 5.89$\cdot10^{20}$\tabularnewline
\bottomrule
\end{tabular}
\par\end{centering}
\caption{{\label{sper-1}Parameters of the linear regressions between the
solar magnetic flux and the mean parameters of the magnetic components:
s is a slope of the regression line, p is two-sided significance of
its deviation from zero, r is a correlation coefficient, and $\sigma$
is the standard error of the estimate. }}
\end{table}

\begin{table}
\begin{centering}
\begin{tabular}{|c>{\centering}p{0.8cm}>{\centering}p{1cm}>{\centering}p{1.5cm}>{\centering}p{1.5cm}>{\centering}p{1.5cm}>{\centering}p{1.5cm}>{\centering}p{1cm}>{\centering}p{1cm}|}
\hline 
\multicolumn{1}{|c|}{} & \multicolumn{1}{>{\centering}p{0.8cm}|}{B-V } & \multicolumn{1}{>{\centering}p{1cm}|}{log${\displaystyle \frac{\overline{B}^{2}}{[G]^{2}}}$ } & \multicolumn{1}{>{\centering}p{1.5cm}|}{${\displaystyle \overline{B^{(T)}}}$,{[}G{]}} & \multicolumn{1}{>{\centering}p{1.5cm}|}{${\displaystyle \overline{B^{(P)}}}$,{[}G{]}} & \multicolumn{1}{>{\centering}p{1.5cm}|}{${\displaystyle \overline{B^{(PA)}}}$,{[}G{]}} & \multicolumn{1}{>{\centering}p{1.5cm}|}{FIP$_{bias}$} & \multicolumn{1}{>{\centering}p{1cm}}{P$_{cyc}$} & Ref\tabularnewline
\hline 
$\chi^{1}$ Ori  & 0.59  & 2.35

2.22

2.60

2.41  & 4.6

4.2

10.2

5.0  & 8.6

5.4

7.8

7.2  & 6.6

4.4

6.8

6.1  & -0.555  & 16.8,L  & {[}1,2,5{]}\tabularnewline
\hline 
EK Dra  & 0.61  & 3.63

3.90  & 22.4

43.8  & 19.2

36.6  & 17.8

34.3  & -0.277  & 9.2,L  & -/-\tabularnewline
\hline 
$\pi^{1}$ Uma  & 0.62  & 2.76  & 7.4  & 9.8  & 8.9  & -0.645  & 2.1,L  & -/-\tabularnewline
\hline 
Sun  & 0.66  & 1.16  & 0.94  & 3.4  & 2.77  & -0.3  & 11  & this work, {[}1,4,5{]}\tabularnewline
\hline 
$\kappa^{1}$Ceti  & 0.68  & 2.88

2.64  & 13.8

11  & 16.9

8.6  & 9.5

7.7  & -0.462  & 5.9,L  & {[}1,2,3,5{]}\tabularnewline
\hline 
$\xi$Boo A  & 0.72  & 3.67  & 10.7  & 29.8  & 9.42  & -0.32  & 4-5, 11,L  & {[}1, 3,5{]}\tabularnewline
\hline 
AB Dor  & 0.86  & 4.92

4.72  & 117.2

80.5  & 244.7

206.2  & 73.1

54.5  & 0.488  & 18,L  & {[}1,3,5{]}\tabularnewline
\hline 
$\epsilon$ Eri  & 0.88  & 2.27

2.08

2.50

2.09

2.63

2.66  & 1.1

2.

2.7

2.1

6.2

3.0  & 13.1

13.2

11.4

9.5

15.3

18.9  & 4.1

10.1

9.7

7.6

10

8.6  & -0.06  & 3,13,L  & {[}1,3,5{]}\tabularnewline
\hline 
$\xi$ Boo B  & 1.16  & 2.60  & 4.9  & 16.5  & 8.1  & -0.18  & 4.3,L  & {[}1,3,5{]}\tabularnewline
\hline 
EV Lac  & 1.36  & 5.54  & 92.3  & 573  & 306  & 0.474  &  & {[}1,3{]}\tabularnewline
\hline 
AD Leo  & 1.54  & 4.76  & 31.3  & 237  & 230  & 0.536  &  & {[}1,3{]}\tabularnewline
\hline 
EQ Peg A  & 1.58  & 5.31  & 131.9  & 423  & 357  & 0.450  &  & {[}1,3{]}\tabularnewline
\hline 
EQ Peg B  & 1.7  & 5.35  & 51  & 468  & 451  & 0.417  &  & {[}1,3{]}\tabularnewline
\hline 
\end{tabular}
\par\end{centering}
\caption{\label{tab:bv}{The B-V is the stellar color index; $\overline{B}$
is the mean magnetic field strength on the stellar surface; $\overline{B^{(T)}}$
is the same for the non-axisymmetric toroidal magnetic field; $\overline{B^{(P)}}$
is the mean strength of the poloidal magnetic field on the stellar
surface (both axisymmetric and nonaxisymmetric); $\overline{B^{(P)}}$
is the same for the axisymmetric poloidal magnetic field; FIP$_{bias}$
is a parameter of the coronal FIP effect; P$_{cyc}$ is period of
the stellar magnetic cycle, L means the long-term magnetic variability
trend (see, \citealt{2017PhDT3E}).} The last column gives list our
references for magnetic data and FIP bias is as follows: {[}1{]}\citet{2015LRSP12.2L};
{[}2{]}\citet{2016AA593A35R}; {[}3{]}\citet{2015MNRAS453.4301S};
{[}4{]}\citet{Brooketal17fip},{[}5{]} \citet{2017PhDT3E}}
\end{table}

\begin{table}
\begin{tabular}{c>{\centering}p{2.6cm}>{\centering}p{2cm}>{\centering}p{2.5cm}>{\centering}p{1.5cm}}
\toprule 
 & Magnetic observations  & References  & FIP$_{bias}$  & References\tabularnewline
\midrule 
$\chi^{1}$Ori  & 2007.1,2008.1,

2010.8,20111.9  & {[}1{]}  & 2001  & {[}5{]}\tabularnewline
\midrule 
EK Dra  & 2007.1,2012.1  & -/-  & 2001  & -/-\tabularnewline
\midrule 
$\pi^{1}$ Uma  & 2007.1  & -/-  & 2001  & -/-\tabularnewline
\midrule 
$\kappa^{1}$Ceti  & 2012.8,

2013.7  & -/-  & 2001  & -/-\tabularnewline
\midrule 
$\xi$ Boo A  & 2013.09  & {\large{}{}{[}2{]}}  & 2007.5  & {[}6,7{]}\tabularnewline
\midrule 
ab Dor  & 2001.12,2002.12  & -/-  & 2001  & {[}8,9{]}\tabularnewline
\midrule 
$\epsilon$ Eri  & 2007.1,2008.1,

2010.1,2011.10,

2012.10,2013.09  & -/-  & 2001.3  & {[}6{]}\tabularnewline
\midrule 
$\xi$ Boo B  & 2013.09  & {\large{}{}{[}3{]}}  & 2007.5  & {[}6,7{]}\tabularnewline
\midrule 
EV Lac  & 2006.8, 2007.7,8/  & {\large{}{}{[}3,4{]}}  & 2001  & {[}8,9{]}\tabularnewline
\midrule 
AD Leo  & 2007.1,2007.2,

2008.1,2008.2  & -/-  & 2001  & {[}9{]}\tabularnewline
\midrule 
EQ Peg A  & 2006.8  & -/-  & 2006.11  & -/-\tabularnewline
\midrule 
EQ Peg B  & 2006.8  & -/-  & 2006.11  & -/-\tabularnewline
\bottomrule
\end{tabular}

\caption{\label{tab:S}Stars and their epochs of the magnetic measurements
and measurements of the FIP -effect. The list our references is as
follows: {[}1{]}\citet{2016AA593A35R}; {[}2{]}\citet{2015MNRAS453.4301S};
{[}3{]}\citet{2016MNRAS1129S}; {[}4{]}\citet{D2-2008MNRAS}; {[}5{]}\citet{2005ApJ622.653T};
{[}6{]}\citet{2006ApJ643.444W}; {[}7{]}\citet{2010ApJ7171279W};
{[}8{]}\citet{2004ApJ617.508T}; {[}9{]}\citet{2008AA491.859L}.}
\end{table}

\subsection{Stellar magnetic observations}

Table \ref{tab:bv} gives the list of the stars included in our analysis.
The set covers a range of spectral classes from the mid G to the mid-M
dwarfs. The color index parameter B-V is taken from the survey of
\citet{2014MNRAS444.3517M}.

Inferring parameters of magnetic activity of the fast rotating stars
became possible after implementation of Zeeman-Doppler Imaging method
(see, \citealp{ZDI2-DnBr-1997AA,2001LNP573.207D,2009ARAA_donat}).
This method is good for determination of the large-scale magnetic
field components of stars with a fast rotation rate. This condition
makes possible to isolate the rotational dynamics of the Zeeman components
in the background spectrum, that is subjected to the Doppler broadening.
All stars in our set are rotating faster than the Sun, for example,
$\epsilon$ Eri and $\xi$ Boo B have the rotational period of ten
days and other stars are rotating faster. They also have the higher
magnetic activity. Interesting that the magnetic activity of late-type
stars is undoubtedly related to the rotational period \citep{1984ApJ...287..769N,Baliunas1995}.

For the fast rotating young suns (having B-V $\sim0.6$) we used results
of \citet{2016AA593A35R} who used archival observational data of
the spectropolarimetric measurements collected in the PolarBase \citep{2014PASP126.469P}.
Results for the K and M-dwarfs are inferred from the data set given
by \citet{2015MNRAS453.4301S}. All magnetic parameters listed in
Table \ref{tab:bv} were calculated using data given in the above-cited
papers. The proxy parameter ${\displaystyle P_{AX}}$ was computed
in following the given data sets. \citet{2016AA593A35R} showed that
typical error for the single measurement of the line-of-sight magnetic
field can be about 50 percents. However, having results for the different
rotational phases the error can be reduced substantially. In comparing
results of inversions from stellar observations with results from
solar magnetic field extrapolation we have to take into account that
the WSO pipeline calculates the PFSS (potential field source surface)
decomposition assuming that the axisymmetric toroidal magnetic field
on the surface is zero, $\overline{B^{(TA)}}=0$. This may be incorrect
\citep{PP14}. Also, this is different from procedure employed for
studying large-scale magnetic field on the fast rotating stars (see,
\citealp{ZDI2-DnBr-1997AA,2001LNP573.207D,2009ARAA_donat,2017MNRAS465L25J}).
Note, that \citet{2016MNRAS4591533V} considered application of the
general procedure, i.e., a combination of the PFSS method and a non-zero
toroidal magnetic field at the surface, for the SDO magnetic field
measurements.

{The Table \ref{tab:bv} contains results about the coronal chemical
composition parameter FIP$_{bias}$ . There we listed parameters given
in review of \citet{2015LRSP12.2L} (hereafter L15). Recently, \citet{Brooketal17fip}(hereafter
B17) re-analyzed the data given in L15 and found that the typical
error of the spectrometric inference of the FIP$_{bias}$ is about
10 to 20 percents. The given data show a correlation between spectral
class and the FIP$_{bias}$. It was found by \citet{2012ApJ753.76W}
and it was discussed in details in L15 and B17. }Number of stars
in our study is smaller because not all the stars, which were analyzed
in L15 and B17, have the magnetic field measurements. Also we have
to take into account that epochs of observations for the magnetic
and chemical composition measurements are very different. The Table
\ref{tab:S} shows epochs of observations for each star. Having data
from different epochs of observations can bias our conclusion. Therefore
analysis of the solar cycle variations of the FIP$_{bias}$ are rather
important for understanding the stellar observations. Some further
details about the solar-cycle variation of the FIP effect and its
relation with the solar magnetic activity were discussed in \citet{piptom18}.

\begin{figure}
\includegraphics[width=0.7\columnwidth]{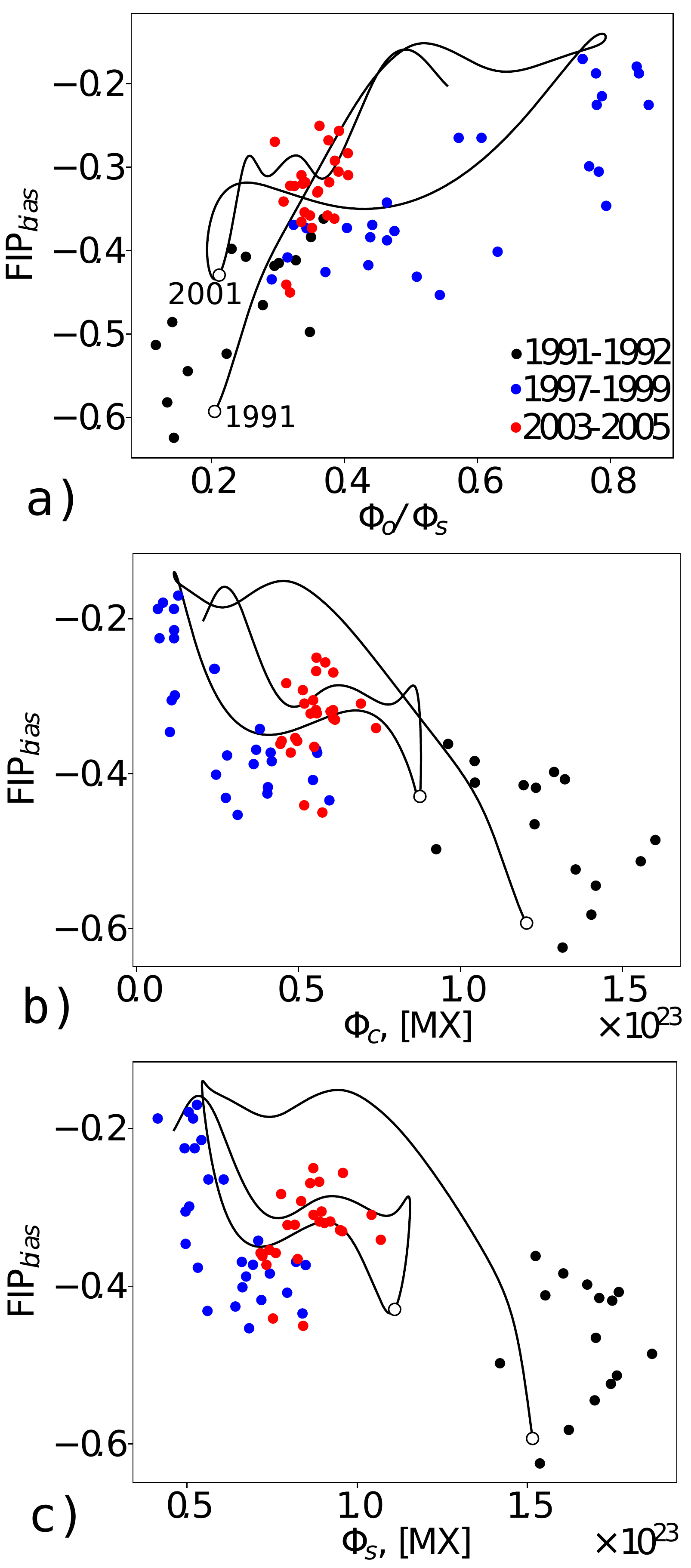}

\caption{\label{fig:fips}{a) Solar cycle relation between the smoothed time-series
of the magnetic fluxes ratio $\Phi_{o}\left(t\right)/\Phi_{s}\left(t\right)$
and the FIP$_{bias}$ (solid line) and point-to-point correlation
between $\Phi_{o}\left(t\right)/\Phi_{s}\left(t\right)$ and the FIP$_{bias}$
for the individual Carrington rotations for the chosen periods of
time (see Figure \ref{fig:sun}b and the main text); b) the same as
(a) for c) the same as (a) for the total closed magnetic flux $\Phi_{c}\left(t\right)$;
c) the same as (s) for the total surface magnetic flux $\Phi_{s}\left(t\right)$.
The years 1991 and 2001 are marked by the white circles.}}
\end{figure}

\section{Results}

\subsection{Solar cycle variation of FIP effect}

{Figures \ref{fig:fips}a,b and c show relations between the smoothed
time-series of $\Phi_{o}/\Phi_{s}$ , $\Phi_{c}$ and $\Phi_{s}$
and the FIP$_{bias}$. All those quantities have the solar cycle variations.
This results in the close relations between evolution curves of the
solar magnetic flux parameters and the FIP$_{bias}$ for the smoothed
time series. The relations in the smoothed time-series do not necessarily
mean the causal connections between the magnetic flux parameters and
the FIP effect. To demonstrate this fact, Figure \ref{fig:fips} shows
correlations between the solar magnetic flux parameters and the FIP
effect for the individual Carrington rotations during three particular
periods of time when the latitude of the spacecraft was near equatorial
regions and it was varying slowly (see, Figure \ref{fig:sun}b). In
this case, the averaged (over one Carrington rotation) parameter FIP$_{bias}$
represent contributions from the equatorial part of the Sun where
the sunspot activity goes on. It is believed that sunspot activity
is the major source of the solar FIP effect. Therefore the obtained
parameter FIP$_{bias}$ could be considered as a better representative
for the integral FIP$_{bias}$ for the Sun as a star than that from
the long-term smoothed time-series. The data sets include the following
time intervals: 1991-1992, 1997-1999 and 2003-2005. The Table\ref{sper}
shows estimations of the Spearman rank correlations and the null hypothesis
probability (that the datasets are uncorrelated). It is found that
the correlation coefficients of $\Phi_{o}/\Phi_{s}$ , $\Phi_{c}$
and the FIP$_{bias}$ have the same sign in all three periods of time.
The magnitude of the correlation coefficients is nearly the same during
the time intervals of 1991-1992 and 1997-1999. It gets smaller during
the time interval of 2003-2005. In all cases the probability of the
null hypothesis is low. }

{Completely different results are found for the correlation coefficients
of the total surface flux $\Phi_{s}$ and the FIP$_{bias}$ . The
correlation coefficients have different signs and magnitude at three
different time intervals. Moreover, the probabilities of the null
hypothesis are higher than for the correlation coefficients of $\Phi_{o}/\Phi_{s}$
, $\Phi_{c}$ and the FIP$_{bias}$, showing that for the period 1991-1992
the total surface flux $\Phi_{s}$ and the FIP$_{bias}$ are likely
uncorrelated. Therefore the relationship between the smoothed time-series
of the parameter $\Phi_{s}$ and the FIP$_{bias}$ is due to their
solar cycle variations. }

{For the unified set of data, that contains points of those three
set, the Spearman rank correlation of the FIP$_{bias}$ and $\Phi_{o}/\Phi_{s}$
is 0.63 with the null hypothesis probability of $10^{-4}$. The same
parameters for the correlation FIP$_{bias}$ and $\Phi_{c}$ are $-0.54$
and $10^{-3}$, respectively. For the unified set of data the correlation
coefficient between $\Phi_{s}$ and the FIP$_{bias}$ is -0.48 and
the null hypothesis probability is $10^{-4}$. As we see before this
does not mean the causal connection between them.}

\begin{table}
\begin{centering}
\begin{tabular}{ccccccc}
\toprule 
 & \multicolumn{2}{c}{1991-1992} & \multicolumn{2}{c}{1997-1999} & \multicolumn{2}{c}{2003-2005}\tabularnewline
\midrule 
 & r  & p  & r  & p  & r  & p\tabularnewline
\midrule 
FIP$_{bias}$ vs{ $\Phi_{o}/\Phi_{s}$ }  & 0.72  & 0.002  & 0.71  & 0.001  & 0.44  & 0.026\tabularnewline
\midrule 
FIP$_{bias}$ vs{ $\Phi_{c}$ }  & -0.578  & 0.023  & -0.72  & 0.001  & -0.276  & 0.181\tabularnewline
\midrule 
FIP$_{bias}$ vs{ $\Phi_{s}$ }  & -0.06  & 0.82  & -0.66  & 0.01  & 0.46  & 0.02\tabularnewline
\bottomrule
\end{tabular}
\par\end{centering}
\caption{{\label{sper}The Spearman's rank correlations, r, and estimations
of the probability of the null hypothesis, p, for the three particular
periods of time. }}
\end{table}

The magnitude of the FIP$_{bias}$ was about of 1.5 factor higher
in maximum of the 22 solar cycle than in the maximum of the 23 cycle.
Results of \citet{Brooketal17fip} shows $\mathrm{FIP}_{bias}\approx-0.6$
near to the maximum of the cycle 24. Following to SIDS sunspot data
center (http://sidc.be/silso) the 24-th solar cycle was considerably
lower than cycles 22 and 23. Therefore the obtained maximum magnitude
of the FIP$_{bias}$ is likely underestimated in comparison with results
of \citet{Brooketal17fip}. {This can be expected because the chosen
procedure of the data reduction does not fully reproduce the full-Sun
measurements of the coronal FIP effect. Preliminary, we conclude that}
the coronal FIP fractionation can be related to the topological properties
of the large-scale magnetic field of the Sun. The connection of the
coronal FIP fractionation with activity cycle is caused by the solar
cycle variations of the topological characteristics of the solar magnetic
fields.

\subsection{Stellar magnetic activity and FIP$_{bias}$}

\begin{figure}
\includegraphics[width=0.8\columnwidth]{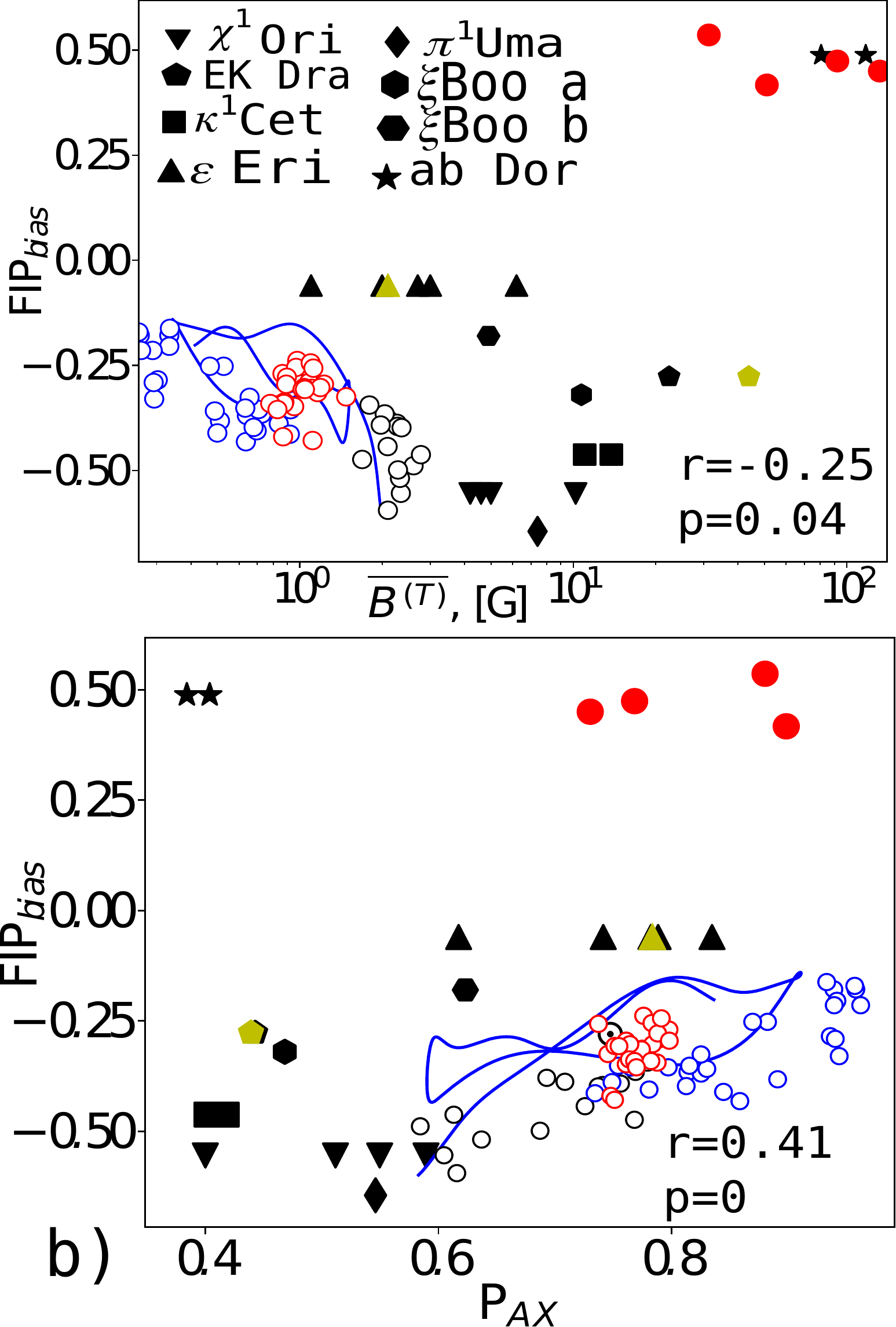}

\caption{\label{fig:FIP-vs} a) The FIP$_{bias}$ vs $\overline{B^{(T)}}$,
the solar-cycle smoothed time series is shown by the solid line, the
data for the individual rotations are shown by circles of the same
color as in Figure \ref{fig:fips}, the red-filled circles show the
M-dwarfs stars, the yellow marks stand for the magnetic observations
which are close to epochs of measurements of FIP$_{bias}$ for stars
$\varepsilon$ Eri and EK Dra (see, the main text), r is the {Spearman
rank correlation} and {p is the probability of the null hypothesis};
b) the same as (a) for the FIP$_{bias}$ vs $P_{AX}$ }
\end{figure}

{Figure \ref{fig:FIP-vs}a and b shows the FIP$_{bias}$ vs the calculated
magnetic parameters $P_{AX}$ and $\overline{B^{(T)}}$ for our sample
of stars including the smoothed solar cycle time-series and data sets
for the individual solar rotations (same as in Figure \ref{fig:fips}).
Variations of the mean strength of the nonaxisymmetric toroidal magnetic
field, $\overline{B^{(T)}}$, show a high level of scattering in the
range from a few Gauss in solar case to hundreds Gauss in case of
AB Dor and the M-dwarfs stars. Also, the solar toroidal magnetic field
is significantly smaller than in the young solar analogs like $\pi^{1}$Uma
or $\chi^{1}$Ori. Those stars have the higher rotation rates and
the higher magnetic activity\citep{Baliunas1995}. The partially convective
solar-type stars have the negative FIP$_{bias}$\citep{2012ApJ753.76W}.
The FIP$_{bias}$ shows some tendency to decrease with the increase
of $\overline{B^{(T)}}$ in direction of the solar cycle variations.
If we take into account the data sets for the individual solar rotations
and restrict ourselves with the case of FIP$_{bias}<0$, the Spearman
rank correlation of the FIP$_{bias}$ vs $\overline{B^{(T)}}$ will
be $r=-0.2$ and the probability of the null hypothesis is $p=0.04$.
The quite high significance level of this correlation is solely because
of including the solar datasets. If we restrict the solar data by
the mean point given in Table \ref{tab:bv}, i.e., FIP$_{bias}=-0.3$
and $\overline{B^{(T)}}\approx1$G, we get $r=-0.03$ and $p=0.89$.
This means that the data are likely uncorrelated. Therefore the results
about the average correlation between FIP$_{bias}$ and $\overline{B^{(T)}}$
are likely not tenable if the magnetic cycle variations are not taken
into account. }

We tried to take into account the magnetic periodicity of some of
the stars from our sample. Using results of \citet{2003AA409.1017M,2013ApJ763L26M}
and \citet{2017PhDT3E} we put the determined periods in the Table
{\ref{tab:bv}}. Comparing the time-series of the magnetic activities
given by \citet{2013ApJ763L26M} we find that epoch of observation
of the coronal chemical composition for the $\epsilon$ Eri, that
is the young K2-dwarf, corresponds to the epoch of the low activity
of the star. In Figures 5 (a) and (b) the points of the magnetic measurement
which is the closest to that phase of the activity of $\epsilon$
Eri is marked by the yellow color. The same conclusion can be drawn
for EK Dra. To determine the approximate phase of the magnetic activity
we use results of \citet{2007AA472.887J} and \citet{2005ApJ622.653T}.
We found that measurements of the FIP$_{bias}$ are related to an
epoch of the relatively low activity of the EK Dra. {Using these
hints and results about the solar cycle variations of the FIP$_{bias}$
we can assume that for the other epochs of the magnetic measurements
of the $\epsilon$ Eri the FIP$_{bias}\le-0.06$ and FIP$_{bias}<-0.3$
for the EK Dra. Taking this into account may reduce the scattering
on the parameter $\overline{B^{(T)}}$in our sample of stars.}

{The AB Dor and the M-dwarfs stars form a group at the up-right corner
of that Figure showing the high level of the toroidal magnetic field
and the positive FIP$_{bias}$. Note that the AB Dor belongs to the
T Tauri branch of cool stars which are on the pre-main sequence stage
of evolution} \citep{2013AA560A.69L}{.}

Figure {\ref{fig:FIP-vs}b shows the FIP$_{bias}$ vs the parameter
$P_{AX}$} (see, Eq\ref{eq:pax}) The{ $P_{AX}$} shows the relative
contribution of the axisymmetric poloidal field on the stellar surface.
On the Sun the given parameter is well correlated with ratio of the
open and closed unsigned magnetic flux $\Phi_{o}/\Phi_{s}$ . {The
the solar-type stars show the decrease of FIP$_{bias}$ with a decrease
of $P_{AX}$}. It is important that the given{ tendency goes in
the same direction as the solar cycle variations of the FIP effect.
If we take into account the data sets for the individual solar rotations
and restrict ourselves with the case of FIP$_{bias}<0$, the Spearman
rank correlation of the FIP$_{bias}$ and $P_{AX}$ will be $r=0.4$
and the probability of the null hypothesis is $p=0$. Excluding the
data for the individual solar rotations, we get $r=0.3$ and $p=0.26$.
Therefore the correlation between parameters of the coronal FIP fractionation
and topology of magnetic activity for the solar-type stars is more
robust than for the correlation of the FIP$_{bias}$ vs $\overline{B^{(T)}}$
. Similarly as above, using arguments of the magnetic periodicity
of the solar-type stars the significance of correlation between the
FIP$_{bias}$ and $P_{AX}$ can be increased.}

The up-right corner of the \ref{fig:FIP-vs}a is occupied by a small
sample of M-dwarf stars. Those stars show a relatively high amplitude
of the poloidal magnetic field (see Table 2). This results in the
high parameter P$_{AX}$ in that sample. The topology of the large-scale
magnetic field of for some of our M-dwarfs (e.g., EQ PegB) resembles
{the Sun's magnetic field during the cycle minims, having very strong
dipole component of the large-scale magnetic field \citep{2008ASPC..384..156D}}.
Those stars have the poloidal magnetic field of strength 1kG and more
\citep{2008ASPC..384..156D,2008MNRAS390.567M}. {It is likely that
the large-scale dynamo on those stars operates in a different way
(for details, see, \citealt{2011MNRAS.418L.133M,2017MNRAS.466.3007P,2017arXiv171201527P}).}
The M-dwarfs show the so-called ``inverse'' FIP$_{bias}$ \citep{2012ApJ753.76W}.
The ``inverse'' FIP$_{bias}$ is also found on the AB Dor which
is a solar analog passing the so-called T-Tauri stage of evolution
\citep{2013AA560A.69L}. Interesting that ratio between the mean magnitudes
of the magnetic field components ( P$_{AX}$) on the AB Dor corresponds
to that of others young solar analogs in our sample. On the other
hand the FIP$_{bias}$ is similar to that found in M-dwarfs and similarly
to M-dwarfs, the AB Dor shows rather a strong mean density of the
poloidal magnetic field.

The exceptional properties of AB Dor are rather interesting. On one
hand, we see that topology of the magnetic field of the star is similar
to the solar-like stars sample. On another hand, the coronal FIP fractionation
on AB Dor likely happens in a way similar to that demonstrated by
the M-dwarfs. From our data (Table \ref{tab:bv}) we also can see
that magnitude of the poloidal magnetic field is growing with increasing
index B-V (associated with the decreasing stellar mass). This is in
agreement with \citet{2008ASPC..384..156D}, \citet{2008MNRAS390.567M}
and \citet{2016MNRAS1129S}. Results of stellar magnetic observations
show that on the M-dwarfs the large-scale poloidal magnetic field
forms the stellar coronal activity. The lack of the observational
data does not allow to discuss a relation between magnetic activity
and the FIP effect on M-dwarfs. Currently, we can only say that the
topological properties of the magnetic activity on M-dwarfs are very
different from the solar-type stars. On another hand results of \citet{2016Natur535.526W}
shows that all low main-sequence stars, including the solar-type stars
and fully convective M-dwarfs, show the same relation of the X-ray
luminosity on the convection Rossby number.

\section{Discussion and conclusions}

Findings of the paper should be considered as preliminary because
of several reasons. There are two major issues in our analysis. Firstly,
the time-averaged FIP effect deduced from the Ulysses dataset can
not fully substitute the Sun as a star FIP-effect determined via spectroscopic
methods (cf, \citealp{Brooketal17fip} ). The cross-calibration between
the in-situ measurements of the solar wind composition and the full-Sun
spectroscopic observations can help to resolve this issue. Secondly,
the introduced parameters of the unsigned open and closed magnetic
flux are found very suitable for studying the solar and stellar wind
\citep{2017MNRAS465L25J}. These parameters may depend on the parameters
of the source surface. Therefore the usability of the correlations
shown in Figure \ref{figmag} for the stellar activity analysis can
be questioned because the radius of the source surface may depend
on both the spectral type and age of a star. Also, the data set of
the our stellar sample is rather small.

Keeping in mind the above issues the main results of the paper can
be formulated as follow. The solar cycle variations of the $\mathrm{FIP}_{bias}$
show a linear relationship to the ratio of the open and closed magnetic
flux of the Sun as a star (see, Figure\ref{fig:fips}a). Taking into
account results of \citealp{Brooketal17fip} we see that the Sun as
a star $\mathrm{FIP}_{bias}$ keeps running in correlation with parameters
$\Phi_{o}/\Phi_{s}$ and $P_{AX}$, varying in the range between $-0.6$
to $-0.2$ from cycle 22 to cycle 24. It is shown that the correlation
between the $\mathrm{FIP}_{bias}$ and $\Phi_{o}/\Phi_{s}$ exists
both for the individual solar rotations and for the solar cycle variation
of the Sun as a star FIP - effect. Also, there is a correlation between
the $\mathrm{FIP}_{bias}$ and $\Phi_{c}$ ( Figure\ref{fig:fips}b).
For the individual solar rotations the statistical significance of
relationship between $\mathrm{FIP}_{bias}$ and $\Phi_{c}$ is worse
than for $\mathrm{FIP}_{bias}$ vs $\Phi_{o}/\Phi_{s}$. However,
our results show that correlation of $\Phi_{c}$ and the FIP$_{bias}$
keep the same sign in all analyzed intervals of the solar cycle. The
closed magnetic flux correlates with the mean toroidal magnetic field
on the solar surface (see, Figure\ref{figmag}b). The large-scale
toroidal magnetic field is often considered as the main source of
the sunspot activity \citep{P55}. \citet{Brooketal17fip} found a
correlation between the Sun as a star FIP-effect and F10.7 radio flux.
The latter is a good proxy for the sunspot activity. We conclude that
the origin of the solar coronal FIP composition depends both on the
topological characteristics of the large-scale solar coronal magnetic
field and the level of the magnetic activity. The parameter $1-\Phi_{o}/\Phi_{s}$
measures the relative contribution of the closed magnetic configurations
to the total unsigned magnetic flux. The parameter $\Phi_{c}$ characterize
the surface-averaged strength of magnetic activity provided by the
closed magnetic loops of the solar active regions. This averaging
may reduce correlation coefficient for the relationship between $\mathrm{FIP}_{bias}$
and $\Phi_{c}$ in compare to $\mathrm{FIP}_{bias}$ and $\Phi_{o}/\Phi_{s}$.
This issue needs to be studied further. Both parameters, the $\Phi_{o}/\Phi_{s}$
and $\Phi_{c}$ vary periodically with the solar cycle. This cause
the solar cycle variation of the Sun as a star FIP - effect.

To investigate the stellar FIP - effect in the same way as the solar
case, we introduce two tracers $P_{AX}$ and $\overline{B^{(T)}}$.
Figure \ref{figmag} and Table \ref{tab:S} show that these tracers
are uniquely related with $\Phi_{o}/\Phi_{s}$ and $\Phi_{c}$. We
employ this in the analysis of the stellar data set. Similar to the
Sun, the solar-type stars show the increase of the stellar $\mathrm{FIP}_{bias}$
goes with the increase of $P_{AX}$ (c.f., $\Phi_{o}/\Phi_{s}$ for
the Sun). Correlation of the $\mathrm{FIP}_{bias}$ to the magnitude
of the toroidal magnetic field, $\overline{B^{(T)}}$ (cf, $\Phi_{c}$
for the Sun) , is weak. This can be a result of the high magnetic
variability in the young solar analogs \citep{Baliunas1995}. The
additional observations of the FIP effect for the different stage
of the stellar magnetic activity can help to clarify the results about
the dependence of the $\mathrm{FIP}_{bias}$ with the magnetic activity
cycles. The relationship between the stellar $\mathrm{FIP}_{bias}$
to the level of the X-ray luminosity of the stellar coronae remains
contradictory. {\citet{2009AIPC1094.796G} found a correlation
between the normalized X-ray luminosity ($L_{\mathrm{X}}/L_{\mathrm{bol}}$)
and FIP on the late-type stars. However analysis of \citet{2012ApJ753.76W},
on the same sample of stars, had shown that this correlation is likely
because of the relationship between the stellar $\mathrm{FIP}_{bias}$
and the spectral class. They did not find a meaningful correlation
of the stellar $\mathrm{FIP}_{bias}$ and the level of the X-ray luminosity
($L_{X}$) of the stellar coronae.} \citet{2003ApJ598.1387P} showed
the high-level significance correlation between the surface magnetic
flux and the X-ray luminosity of the low main-sequence stars. The
large-scale organization of the magnetic activity changes with age
and with the spectral class of a star \citep{2008ASPC..384..156D,2009AA496.787R,2016MNRAS1129S}.
Summarizing the above facts, we confirm our conclusions that in the
solar analogs the magnitude of the $\mathrm{FIP}_{bias}$ is connected
both with the magnetic variability of stars and topology of the coronal
magnetic field. 

The above conclusion can be suitable for the theoretical scenario
of the solar FIP effect which was developed by \citet{2015LRSP12.2L}(hereafter
L15). Comparing to other ideas, e.g., \citet{1999ApJ521.859S} (also,
see the review of \citealp{tom13}), the scenario of L15 has an advantage
explaining both the normal (solar-like) and the inversed FIP effect
from the same mechanism. The key idea behind his model is that the
FIP fractionation is induced by the so-called ponderomotive force.
This force results from propagation of resonant alfvenic waves along
the closed magnetic field lines. Direction and amplitude of the ponderomotive
force depend on the gradient of wave energy in the low corona. For
the resonant waves the energy grows up and this results in the upward
ponderomotive force acting on ions of the low FIP. In the open magnetic
field structures and for the off-resonant waves the wave energy decreases
in the major part of the low corona (see \citealp{2004ApJ614.1063L}).
Strong reflections of waves from upper boundary of transition region
results in inversion of the ponderomotive force and inversion of the
FIP effect. The detailed calculations made by \citet{2012ApJ753.76W}
and Laming(2015) demonstrated the efficiency of the given mechanism
both for the Sun and the M-dwarfs stars. Note, that the solar relationship
between the $\mathrm{FIP}_{bias}$ and $\Phi_{o}/\Phi_{s}$ can be
also fitted in the model of \citet{1999ApJ521.859S}. They suggested
that FIP fractionation is developed due to the wave heating of minor
ions in on the closed configurations of large coronal loops. However,
their theory does not explain the inversed FIP effect.

The given scenario gives a natural explanation for the solar cycle
relation between parameters of the open and close magnetic flux and
the coronal $\mathrm{FIP}_{bias}$ for the Sun as a star. Note that
the alfvenic waves are one of the possible mechanism for the solar
corona heating \citep{2005pscibookA}. Interesting that estimations
of L15 showed that magnitude of the density energy of alfvenic waves
is the one that it is necessary for to maintain the magnitude of the
$\mathrm{FIP}_{bias}$ in the magnetic arcade. This density energy
of alfvenic waves corresponds to that needed for the coronal heating,
as well. The mechanism suggested by L15 was never applied to estimate
the $\mathrm{FIP}_{bias}$ for the Sun as a star. Also, the linear
estimations of the ponderomotive force profiles given by \citet{2015LRSP12.2L}
can be inconsistent as we have to take into account the nonlinear
effects (see,\citealp{2017ApJ840.64S}).

\citet{2004ApJ614.1063L} and \citet{2012ApJ753.76W} argued that
the inverse $\mathrm{FIP}_{bias}$ on M-dwarfs can be related with
strong reflections of alfvenic waves from the upper boundary of the
transition region. The effect of strong reflections can be caused
by difference of the vertical scales of transition region at the base
of the opposite legs of the magnetic arcades (see, \citealp{2004ApJ614.1063L}
and \citealp{1984ApJ277.392H}). This is expected to find out on active
M-dwarfs. Those stars show magnetic field organization which consists
of the large-scale patches of strong magnetic field \citep{2008ASPC..384..156D}. 

Roughly speaking the inverse FIP$_{bias}$ on M-dwarfs can be related
to the dramatic increase of magnitude of the coronal magnetic field.
The given mechanisms are considered as the theoretical scenarios which
need the further development.

Our conclusions can be summarized as follows. The solar-like FIP fractionation
of the coronal elements is likely related to the topology of the large-scale
coronal magnetic field and activity level of the toroidal magnetic
field. The $\mathrm{FIP}_{bias}$of the Sun as a star varies in the
solar cycle in inverse proportion to the contribution of the unsigned
closed magnetic flux, which depends on ratio the mean strength of
the large-scale poloidal and toroidal magnetic field on the solar
surface. Available theoretical scenarios suggest that the solar-like
$\mathrm{FIP}_{bias}$ as well as the inverse $\mathrm{FIP}_{bias}$
can be described in the unified picture relating effects of the large-scale
the topology of magnetic fields and the nonlinear processes of coronal
waves propagation.

\textbf{Acknowledgments.} This work was conducted as a part of FR
program II.16 of ISTP SB RAS. VP thanks the financial support from
of RFBR grant 17-52-53203. 
\begin{center}
\textbf{REFERENCES}  \bibliographystyle{elsarticle-harv}

\par\end{center}
\end{document}